\documentstyle[11pt]{article}
\def\beq{\begin{equation}}
\def\eeq{\end{equation}}
\def\beqa{\begin{eqnarray}}
\def\eeqa{\end{eqnarray}}
\begin{document}
\begin{titlepage}
\vspace*{-1cm}
\noindent
\phantom{bla}
\hfill{UMHEP-390}
\\
\vskip 2.5cm
\begin{center}
{\Large\bf Anatomy of a Weak Matrix Element}
\end{center}
\vskip 1.5cm
\begin{center}
{\large John F. Donoghue and Eugene Golowich}
\\
\vskip .3cm
Department of Physics and Astronomy \\
University of Massachusetts \\
Amherst MA 01003 USA\\
\vskip .3cm
\end{center}
\vskip 2cm
\begin{abstract}
\noindent
Although the weak nonleptonic amplitudes of the Standard Model
are notoriously difficult to calculate, we have produced a modified
weak matrix element which can be analyzed using reliable methods.
This hypothetical nonleptonic matrix element is expressible in terms
of the isovector vector and axialvector spectral functions
$\rho_{\rm V}(s)$ and $\rho_{\rm A}(s)$, which can be determined
in terms of data from tau lepton decay and $e^+ e^-$ annihilation.
Chiral symmetry and the operator product expansion are used to constrain
the spectral functions respectively in the low energy and the high energy
limits.  The magnitude of the matrix element thus determined is compared
with its `vacuum saturation' estimate, and in the future may be accessible
with lattice calculations.
\end{abstract}
\vfill
\end{titlepage}
\vskip2truecm
\section{\bf Introduction}

To determine the origin of the $\Delta I = 1/2$ rule
remains one of the unsolved problems of the Standard Model.$^{\cite{0}}$
Despite years of effort involving theoretical techniques ranging
from quark model calculations to lattice-gauge studies, a
quantitative understanding of the $\Delta I = 1/2$ enhancement
is still lacking.  The reason for this can be seen from
the structure of the $\Delta S = 1$ nonleptonic weak hamiltonian,
\beq
{\cal H}_{{\rm wk}}^{\Delta S=1} = {g_2^2 \over 8} V^*_{\rm ed}V_{\rm us}
\int d^4 x ~iD_{\mu\nu}(x,M_{\rm W})~
T( J^{\mu\dagger}_{1+i2} (x) J^\nu_{4+15} (0)) \ \ ,
\label{0}
\eeq
where $iD^{\mu\nu}(x,M_{\rm W})$ is the $W$--boson propagator,
$g_2$ is the gauge coupling strength for weak isospin and
$J^\mu_k$ are the left--handed quark weak currents,
\beq
J^\mu_k = {\bar q}{\lambda_k \over 2}\gamma^\mu (1 + \gamma_5) q
\ \ ,
\label{1}
\eeq
where $q = (u~d~s)$.  Strong interaction effects are present in Eq.~(\ref{0})
at all energy scales up to the $W$-boson mass.  It is this fact which
signals the difficulty of dynamically reproducing the $\Delta I = 1/2$ rule.

However, experience from both experiment and theory has
taught us much about low-energy $QCD$.  It is possible
that a hybrid approach which combines data with theory
might provide insight as to the source of the $\Delta I = 1/2$ effect.
The calculation to follow will constitute a first step in that
direction.  It involves just the vector-vector part of the
operator in Eq.~(\ref{0}) (note that we drop the $KM$ factors)
\beq
{\cal H}^{\Delta S=1}_{\rm vv}
= {g_2^2\over 8} \int d^4 x ~iD_{\mu\nu}(x,M_{\rm W})~
T\left( V_{1-i2}^\mu (x) V_{4+i5}^\nu (0)\right)\ \ .
\label{2}
\eeq
and its matrix elements taken between single-particle
pseudoscalar meson states,
\beqa
{\cal M}_{K^-\to \pi^-} &=& \langle \pi^- |{\cal H}^{\Delta S=1}_{\rm vv}|
K^- \rangle \ \ , \nonumber \\
{\cal M}_{{\bar K}^0\to \pi^0} &=& \langle \pi^0
|{\cal H}^{\Delta S=1}_{\rm vv}|{\bar K}^0 \rangle \ \ .
\label{2a}
\eeqa
Our goal is to determine the matrix elements
${\cal M}_{{\bar K}\pi}$ as accurately as possible,
despite the complications of the strong interactions,

The main theoretical tool will be chiral symmetry.  It is easy to show
that the single-meson matrix elements of the chiral operator of
Eq.~(\ref{0}) vanish in the soft meson limit.  By contrast,
the ${\cal M}_{K\pi}$ of Eq.~(\ref{2a}) have a nontrivial form in the
chiral limit because they involve a {\it vector-vector}
operator.$^{\cite{1}}$   By working in the soft meson limit, we can
utilize a spectral representation to study the matrix elements.
Constraints on the low-energy and high-energy limits of the spectral
integrals are obtainable from chiral symmetry and $QCD$ sum rule methods,
and data can be used to fill in much of the rest.  It is in this sense
that the `anatomy' of these matrix elements are revealed.

Although the ${\cal M}_{K\pi}$ are thus
a kind of theoretical laboratory from which we can
learn something of value, they themselves do not
reproduce the $\Delta I = 1/2$ signal.  This can best be seen
in terms of an effective lagrangian description.  Under the
usual classification scheme of $SU(3)_L\times SU(3)_R$,
the operator of Eq.~(\ref{2}) transforms as a
component of an $({\bf 8},{\bf 8})$ family of (parity--conserving)
operators.$^{\cite{2}}$  An $SU(3)$ invariant lagrangian which can
characterize the momentum independent parts of the transition amplitudes is
\beq
{\cal L}^{(8,8)}_{ab} = g {\rm Tr}~ \left( \lambda_a \Sigma \lambda_b
\Sigma^\dagger
+ \lambda_b \Sigma \lambda_a \Sigma^\dagger \right)~,
\qquad (a,b =1,\dots 8) \ \ ,
\label{2b}
\eeq
where $g$ is a constant, $\Sigma$ represents the meson fields,
\beq
\Sigma \equiv \exp \left( i \phi_c \lambda_c /F \right)~, \qquad
(c = 1,\dots 8) \ \ ,
\label{2c}
\eeq
and $F$ is the meson decay constant.  For the nonleptonic weak
$({\bf 8},{\bf 8})$ operator, the ratio of $\Delta I = 1/2$ and
$\Delta I = 3/2$ amplitudes is fixed.  Thus, given the isospin decomposition
\beqa
{\cal M}_{K^-\to \pi^-} &=& {\cal M}_{1/2} + {\cal M}_{3/2}~, \nonumber \\
{\cal M}_{{\bar K}^0\to \pi^0} &=& -{1\over \sqrt{2}}{\cal M}_{1/2}
+ \sqrt{2}{\cal M}_{3/2}\ \ ,
\label{2d}
\eeqa
it is straightforward to show
\beq
{{\cal M}_{3/2} \over {\cal M}_{1/2}} = -{2\over 5}\ \ .
\label{2e}
\eeq
\section{\bf Derivation of a New Chiral Sum Rule}

In the following, we shall express
the ${\bar K}\to \pi$ transition amplitudes in the form of
a spectral representation.  {\it Throughout, all our calculations will
be performed in the chiral limit of massless $u,d,s$ quarks}.  The first
step in the process is to invoke the current algebra results$^{\cite{3}}$
\beqa
\lefteqn{\lim_{p\to 0} \langle \pi^-_{\bf p} |
T\left( V^\mu_{1-i2} (x) V^\nu_{4+i5} (0)\right) | K^-_{\bf p} \rangle = }
\nonumber \\
& & -{1\over F^2}\langle 0| T\left( V^\mu_3 (x)
V^\nu_3 (0) - A^\mu_3 (x) A^\nu_3 (0)\right) | 0 \rangle \ ,
\label{2h}
\eeqa
and
\beqa
\lefteqn{\lim_{p\to 0} \langle \pi^0_{\bf p}
| T\left( V^\mu_{1-i2} (x) V^\nu_{4+i5} (0)\right) | {\bar K}^0_{\bf p}
\rangle = } \nonumber \\
& & {3\over \sqrt{2}F^2}\langle 0| T\left( V^\mu_3 (x)
V^\nu_3 (0) - A^\mu_3 (x) A^\nu_3 (0)\right) | 0 \rangle \ ,
\label{2i}
\eeqa
where we take the vacuum to be an $SU(3)$ singlet.
Note the dependence upon the
difference of vector and axialvector terms.  These quantities
can be expressed in terms of spin-one spectral functions
$\rho_{\rm V,A}(s)$ as defined by$^{\cite{4}}$
\beqa
\lefteqn{\langle 0|T\left( V^\mu_a (x) V^\nu_b (0)\right)
|0\rangle =} \nonumber \\
& & i\delta_{ab} \int_0^\infty ds~ \rho_{\rm V} (s)~(-sg^{\mu\nu}
- \partial^\mu \partial^\nu) \int {d^4p\over (2\pi)^4}~{e^{-ip\cdot x}\over
p^2 - s + i\epsilon}
\label{4a}
\eeqa
and
\beqa
\lefteqn{\langle 0|T\left( A^\mu_a (x) A^\nu_b (0)\right)|0\rangle =
- i\delta_{ab}F_\pi^2 \partial^\mu\partial^\nu
\int {d^4p\over (2\pi)^4}~{e^{-ip\cdot x}\over
p^2 + i\epsilon}}  \nonumber \\
& &  + i\delta_{ab}
\int_0^\infty ds ~\rho_{\rm A} (s)(-sg^{\mu\nu} -
\partial^\mu\partial^\nu) \int {d^4p\over (2\pi)^4}~{e^{-ip\cdot x}\over
p^2 - s + i\epsilon} \ .\label{4b}
\eeqa
Note that in the chiral limit, the spin $0$ contribution is given entirely
by the pion pole.  It is thus possible to write the
${\cal M}_{{\bar K}\pi}$ in spectral form,
\beq
{\cal M}_{K^-\to \pi^-} = {3i\over 32\sqrt{2}
\pi^2}{G_\mu \over F^2} {\cal A}
\qquad {\rm and} \qquad
{\cal M}_{{\bar K}^0\to \pi^0} = -
{9i\over 64\pi^2}{G_\mu \over F^2} {\cal A}\ \ ,
\label{4c}
\eeq
where
\begin{equation}
{\cal A} = M_{\rm W}^2
\int_0^\infty ds ~s^2 \ln \left({s\over M_{\rm W}^2}\right)
{\rho_{\rm V}(s) - \rho_{\rm A}(s) \over s - M_{\rm W}^2 + i\epsilon}
\ \ .
\label{5}
\end{equation}

\vspace{1.0 cm}

The combination of vector and axialvector spectral functions
in Eq.~(\ref{5}) is reminiscent of similar forms appearing in
other well-known sum rules {\it viz.},
\begin{eqnarray}
\int_{0}^\infty ds ~{\rho_{\rm V}(s) - \rho_{\rm A}(s)\over s}
&=& -4{\bar L}_{10}\ \ , \label{6} \\
\int_{0}^\infty ds ~(\rho_{\rm V}(s) - \rho_{\rm A}(s)) &=& F_\pi^2 \ \ ,
\label{7} \\
\int_{0}^\infty ds \ s~(\rho_{\rm V}(s)
- \rho_{\rm A}(s)) &=& 0 \ \ , \label{8} \\
\int_{0}^\infty ds ~s\ln\left({s\over\Lambda^2}\right)~
(\rho_{\rm V}(s) - \rho_{\rm A}(s)) &=&
-{16\pi^2 F_\pi^2\over 3e^2}  (m^2_{\pi^\pm} - m^2_{\pi^0} )
\ \ .  \label{9}
\end{eqnarray}
In the first sum rule, ${\bar L}_{10}$ is related to
the renormalized coefficient $L_{10}^{(r)}(\mu)$ of
an ${\cal O}(E^4)$ operator in the effective chiral lagrangian
of $QCD$$^{\cite{5}}$,
\beq
{\bar L}_{10} = L_{10}^{(r)}(\mu)  + {144\over \pi^2}
\left[ \ln \left( {m_\pi^2\over \mu^2}\right) + 1\right]
\simeq -6.84\times 10^{-3} \ \ .
\label{9a}
\eeq
The next two relations are respectively the first
and second Weinberg sum rules.$^{\cite{6}}$  Finally,
Eq.~(\ref{9}) is the formula for the $\pi^\pm$-$\pi^0$ mass splitting
which was derived long ago using soft-pion methods.$^{\cite{7}}$  Although
containing an arbitrary energy scale $\Lambda$, this last
expression is actually independent of $\Lambda$
by virtue of the second Weinberg sum rule.  As one would expect,
Eq.~(\ref{5}) reduces to the structure of Eq.~(\ref{9}) upon dropping
the prefactor of $M_{\rm W}^2$ and taking the limit
$M_{\rm W}^2 \to 0$.  Although it is the expression in Eq.~(\ref{5})
which is central
to our analysis, we shall see that these other sum rules serve as
crucial checks on the reliability of our determination.
\section{\bf The Spectral Functions}
As mentioned earlier, the difficulty in calculating the weak nonleptonic
matrix element lies in the need to understand all the physics between
$s = 0$ and $s = M_{\rm W}^2$.  Fortunately, with the particular
combination found in Eq.~(\ref{5}) ({\it i.e.} $\rho_{\rm V} - \rho_{\rm A}$),
we are able to overcome this hurdle.  We have recently provided a
detailed phenomenological overview$^{\cite{dg}}$ of
$\rho_{\rm V}$-$\rho_{\rm A}$ and of the chiral sum rules in
Eqs.~(\ref{6}-\ref{9}).  Briefly, the important ingredients are as follows.

At very low energy, the difference of spectral functions is uniquely
determined by chiral symmetry to be$^{\cite{gl}}$
\beq
\rho_{\rm V}(s) - \rho_{\rm A}(s) \sim
{1\over 48\pi^2} \left( 1 - {4m_\pi^2\over s} \right)^{3/2}
\theta(s - 4m_\pi^2) + {\cal O}(p^2) \ \ .
\label{10}
\eeq
At somewhat higher energies, the vector spectral function $\rho_{\rm V}$
may be determined from both $e^+ e^-$ annihilations and decay of the tau
lepton.$^{\cite{ts}}$  As a consistency check, the two data sets have been
shown to give compatible information in the $\rho (770)$ resonance
region.$^{\cite{gan}}$  The data reveals that as energy is increased above
threshold, $\rho_{\rm V}$ is dominated first by $2\pi$ and then $4\pi$
resonances.  At even higher energy, multipion production leads to a continuum
component which ultimately approaches the asymptotic $QCD$ prediction,
\beq
\rho_{\rm V} (s) \sim {1\over 8\pi^2}\left( 1 + {\alpha_s (s)\over\pi}
\right)\ .
\eeq
For the axialvector spectral function $\rho_{\rm A}$, things
proceed similarly, except that one must rely solely upon tau decay
data.  Also, above threshold it is the $3\pi$ sector which
first occurs and the dominant resonance contribution
is that of $a_1 (1260)$.  In the large energy limit, $QCD$ predicts
that $\rho_{\rm A}$~=~$\rho_{\rm V}$ to all orders in perturbation
theory.$^{\cite{rem}}$  The {\it difference} between
$\rho_{\rm V}$ and $\rho_{\rm A}$ arises from non-perturbative
effects and may be estimated by using the operator product
expansion.$^{\cite{rus}}$  The asymptotic energy dependence behaves
as $s^{-3}$, and in the approximation of vacuum saturation we find
\beq
\rho_{\rm V}(s) - \rho_{\rm A}(s) \sim
{8\over 9}
{\alpha_s \langle \sqrt{\alpha_s }
{\bar q}q\rangle_0^2 \over s^3} \simeq
{3.4\times 10^{-5}\ {\rm GeV}^6
\over s^3}\ \ ,
\label{13}
\eeq
where we take $\alpha_s (5~{\rm GeV^2}) \simeq 0.2$.
In view of the tiny numerator, this asymptotic tail provides
only a very small contribution to the sum rule integrals and
thus its precise value is not very important.

A phenomenological analysis of the chiral sum rules has been
performed in Ref.~\cite{dg}.  Both numerical and analytical
approaches have been employed to convert the empirical knowledge
of $\rho_{\rm V} - \rho_{\rm A}$ into statements about the sum rules.
Figure~1 displays a typical solution found there for
$\rho_{\rm V}$~-~$\rho_{\rm A}$.  It represents a fit of the $\tau \to 2\pi$,
$3\pi$, $4\pi$ decay spectra and branching ratios and $e^+ e^- \to
2\pi$, $4\pi$ cross sections.  Other solutions are found to have a
highly similar appearance, and we refer the reader to Ref.~\cite{dg}
for details.  As a whole, the solution set has the correct high energy
and low energy behavior, as well as reproducing the correct values of
the four chiral sum rules listed in Eqs. (17)-(20).  Of course, the
phenomenology is based on existing data with attendant uncertainties,
especially with respect to the precise values of the $3\pi$ and
$4\pi$ branching ratios in tau decay.  The form of the curve
in Fig.~1 is expected to be subject to minor refinements as future
experimental results become available.  However, such a determination
of the spectral functions allows us to calculate the
weak matrix element ${\cal A}$, and we ascertain its value to be
\beq
{\cal A} \ = \ -0.062 \pm 0.017 ~{\rm GeV}^6  \ \ .
\label{w4}
\eeq
The error bars we assign about the central value is our estimate
of the uncertainties associated with the current data base.
We find that the most important contributions to the dispersive integral in
Eq.~(\ref{5}) arise from the `intermediate' energy region,
$1<s ({\rm GeV}^2)<10$.  This is seen to be a result of the factor
of $s^2$ in the integrand which suppresses the effects at low $s$ and
of the vanishing of $\rho_{\rm V}$~-~$\rho_{\rm A}$ as $s \to \infty$
which controls the high energy region.

We conclude this section with a model-dependent estimate of the
amplitude ${\cal A}$.  Since rigorous calculation of nonleptonic
matrix elements has traditionally not been available, various
approximation schemes have instead been employed, among them quark
models and the vacuum saturation approach.  For definiteness, we shall
adopt the latter approach here.  The short distance expansion
\beq
T\left( V_{1-i2}^\mu (x) V_{4+i5}^\nu (0)\right) =
V_{1-i2}^\mu (0) V_{4+i5}^\nu (0) \ + \ \ {\cal O}(x) \ \ ,
\label{sd1}
\eeq
along with the evaluation
\beq
\int d^4x ~iD_{\mu\nu}(x,M_{\rm W}) = {ig_{\mu\nu}\over M_{\rm W}^2}\ \ ,
\label{sd2}
\eeq
allows us to write
\beq
{\cal M}_{{\bar K}\pi} \simeq -{iG_\mu \over \sqrt{2}}
\langle \pi |{\bar d}_i \gamma^\mu u_i {\bar u}_j \gamma_\mu s_j
|{\bar K} \rangle \ \ .
\label{sd3}
\eeq
In the vacuum saturation approximation, Eq.~(\ref{sd3})
is determined by performing a Fierz transformation
and inserting the vacuum intermediate states.  Remembering that
we are working in the chiral limit, we find
\beq
{\cal A}_{\rm vac} = -{ 32\pi^2\over 9} \langle {\bar u} u\rangle_0^2
\simeq - 0.033~{\rm GeV}^6 \ \ .
\label{vac}
\eeq
This is about half the spectral evaluation given above.
\section{\bf Concluding Remarks}To our knowledge, this is the
first time that anyone has undertaken evaluation of a weak nonleptonic
matrix element using a battery of methods which, at least in principle,
are both theoretically sound and practical to apply.  The resulting
analytic control allows us to be able to identify the most important
of the underlying physics.  We have found that the dominant contributions
to the matrix element come from the region of intermediate energies,
which is the one over which theorists
have the least control.  That is, at the lowest energies, rigorous
predictions can be made using chiral symmetry, while at the
highest energies methods of perturbative and nonperturbative $QCD$ may
be invoked.  However, the intermediate energy region continues to resist
the development of reliable analytic methods.  Fortunately, the
existence of a data base allows this range to be estimated.   We have
done the best that can be accomplished at this time.  As the quality
and quantity of data improves, we expect to be able to reduce accordingly
the uncertainty in our determination.

It will be interesting to compare our evaluation of
${\cal M}_{{\bar K} \to \pi}$ with those of computer simulations
in lattice gauge theory.  It could well be that our
phenomenological determination, even with the presence of the
aforementioned uncertainties, is still more accurate
than those of present day lattice studies.  The energy
range $1 < s ({\rm GeV}^2) < 10$ is a problematic one for
lattice simulations because the energy scale is comparable to the
inverse lattice size used in current computations.  Lattice
artifacts may then be significant.  In addition, although
today's lattice studies are becoming more accurate in the prediction of
resonance masses and couplings, they are not yet competitive in
matrix element analysis with the experimental determinations which
were the prime ingredient of our evaluation.

The weak matrix element that we have studied does not
immediately provide the key to solving the riddle of
the $\Delta I = 1/2$ rule.  However, it does provide an
interesting theoretical laboratory for the development
of improved calculational methods which may prove to be
efficient in addressing this obdurate puzzle.  Additional work
on this approach is underway and will be reported on in future
publications.

The research described in this paper was supported in part by
the National Science Foundation.

\eject
\begin{center}{\large{\bf Figure Caption}}
\end{center}
\vspace{0.3cm}
\begin{flushleft}
Fig.~1 \hspace{0.2cm} The spectral function $\rho_{\rm V} - \rho_{\rm A}$
\end{flushleft}
\vfill \eject

\begin{thebibliography}{99}
\bibitem{3} Note that these formulae imply the result in Eq.~(\ref{2e}).
\bibitem {rem} We remind the reader that we are working with massless
$u$, $d$ quarks.
\end{thebibliography}
\end{document}